\documentclass[journal=jacsat,manuscript=article]{achemso}
\setkeys{acs}{email = false}
\usepackage[version=3]{mhchem}

\author{Jonas D. Ziegler*}
\affiliation{Photonics Laboratory, ETH Zürich, CH-8093 Zürich, Switzerland}
\author{Sotirios Papadopoulos}
\affiliation{Photonics Laboratory, ETH Zürich, CH-8093 Zürich, Switzerland}
\alsoaffiliation{Université de Strasbourg, CNRS, Institut de Physique et Chimie des Matériaux de Strasbourg, UMR 7504, Strasbourg, France}
\author{Antti J. Moilanen}
\affiliation{Photonics Laboratory, ETH Zürich, CH-8093 Zürich, Switzerland}
\author{Marcelo M. Valenzuela}
\affiliation{Photonics Laboratory, ETH Zürich, CH-8093 Zürich, Switzerland}
\author{Qia Lin}
\affiliation{Photonics Laboratory, ETH Zürich, CH-8093 Zürich, Switzerland}
\author{Kseniia Mosina}
\affiliation{Department of Inorganic Chemistry, University of Chemistry and Technology Prague, Prague, Czech Republic}
\author{Takashi Taniguchi}
\affiliation{International Center for Materials Nanoarchitectonics,  National Institute for Materials Science, Tsukuba, Ibaraki 305-004, Japan}
\author{Kenji Watanabe}
\affiliation{Research Center for Functional Materials, National Institute for Materials Science, Tsukuba, Ibaraki 305-004, Japan}
\author{Zdenek Sofer}
\affiliation{Department of Inorganic Chemistry, University of Chemistry and Technology Prague, Prague, Czech Republic}
\author{Florian Dirnberger}
\affiliation{Zentrum für Quantum Engineering (ZQE), Technical University of Munich, Garching, Germany}
\alsoaffiliation{Physics Department, TUM School of Natural Sciences, Technical University of Munich, Munich, Germany}
\alsoaffiliation{Munich Center for Quantum Science and Technology (MCQST), Technical University of Munich, Garching, Germany.}
\author{Lukas Novotny}

\affiliation{Photonics Laboratory, ETH Zürich, CH-8093 Zürich, Switzerland}

\title[Electrical excitation of a 2D magnet]
{All-electrical near-field injection of excitons in a van der Waals antiferromagnet}

\abbreviations{2D, CrSBr, PL, EL, AFM}

\begin{document}
\newpage
\begin{abstract}Van der Waals materials have become a promising building block for future electronics and photonics. The two-dimensional magnet CrSBr came into the spotlight of solid state research due to its intriguing combination of antiferromagnetic order, strong light-matter coupling and unusual quasi-1D electronic bandstructure. This study reports the electrical excitation of excitons in CrSBr layers from cryogenic temperatures up to room temperature. By exploiting the energy transfer via tunneling electrons in a graphene tunnel junction  strongly bound excitons are excited in proximate CrSBr layers. This facilitates electrically-excited emission from CrSBr crystals ranging in thickness from a bilayer up to 250\,nm, in which the strong linear polarization of the electroluminescence confirms the excitonic origin. For thicker layers, clear evidence for the electrically excited emission from self-hybridized exciton polaritons is observed, highlighting the strong coupling between optical excitations and confined photon modes in CrSBr. These results pave the way for future applications in spintronic and optical readout of magnetic properties.
\end{abstract}
\section{Introduction}
Two-dimensional (2D) materials have emerged at the forefront of condensed matter research, providing a versatile platform to explore novel quantum phenomena and enabling potential applications in next-generation devices. Within this realm, van der Waals (vdW) magnets are particularly attractive, as their weak interlayer interactions facilitate precise atomic-scale confinement down to the monolayer limit \cite{Gong2017,Huang2017, Klein2018}. In addition to metallic and semimetallic vdW materials, semiconducting vdW magnets have attracted considerable attention owing to their tunable optical and electronic properties, making them promising candidates for fundamental research \cite{Kang2020,Hwangbo2021,Dirnberger2022, Wu2019} and spintronic applications \cite{Song2018, BoixConstant2024}.

Among these materials, chromium sulfur bromide (CrSBr) has rapidly gained prominence due to its compelling combination of desirable properties \cite{Klein2023a, Ziebel2024,Meineke2024,Lin2024,Liebich2025}. CrSBr exhibits a relatively high Néel temperature \cite{Wilson2021}, air stability, and strongly bound excitons with large oscillator strengths \cite{Ruta2023}. Furthermore, its intrinsic antiferromagnetic (AFM) ordering and strong electronic anisotropy render it particularly suitable for applications in quantum technologies and spintronics \cite{Jungwirth2018,Chen2024}.

A particularly intriguing characteristic of CrSBr lies in the interplay between its optical excitations and magnetic ordering. The strong oscillator strength in CrSBr allows for the formation of self-hybridized exciton polaritons, enabling unprecedented control and optical tunability of excitonic states \cite{Dirnberger2023}. These unique light-matter interactions provide new opportunities to manipulate quantum states through optical methods.

However, effectively exploiting these optical and spintronic features for practical device applications requires electrical excitation and control of excitonic states. Traditionally, electrical excitation in conventional 2D semiconductors is achieved through separate injection of electrons and holes \cite{Bajoni2008}, a mechanism critical for realizing efficient spin-based devices and coherent quantum control.

In this work, we introduce a novel electrical excitation mechanism in CrSBr, mediated by the coupling of tunneling electrons to excitons, that occurs independently of flake thickness. Specifically, we demonstrate for the first time electrically excited excitonic states in CrSBr flakes ranging from bilayer (2\,nm) thickness up to bulk crystals (250\,nm). Notably, thicker samples exhibit multiple resonances consistent with previously observed self-hybridized exciton polaritons \cite{Dirnberger2023, Canales2021}. This near-field excitation mechanism, along with the unconventional magnetic response of the material, opens exciting pathways for the manipulation of excitonic and electronic states in magnetic vdW materials.

\section{Results and discussion}

\subsection{Energy transfer excitation from tunneling electrons}
We employ a recently reported near-field excitation technique driven by tunneling electrons for CrSBr \cite{Papadopoulos2022, Wang2023}. In this design, the active material is positioned outside the direct electrical conduction pathway. This forms an open electrode configuration, similar to open cavity designs. Figure 1(a) illustrates the device design, which consists of a CrSBr layer of variable thickness placed on top of a graphene-hexagonal boron nitride (hBN)-gold tunnel junction.

\begin{figure*}[h]
	\centering
			\includegraphics[scale=0.5]{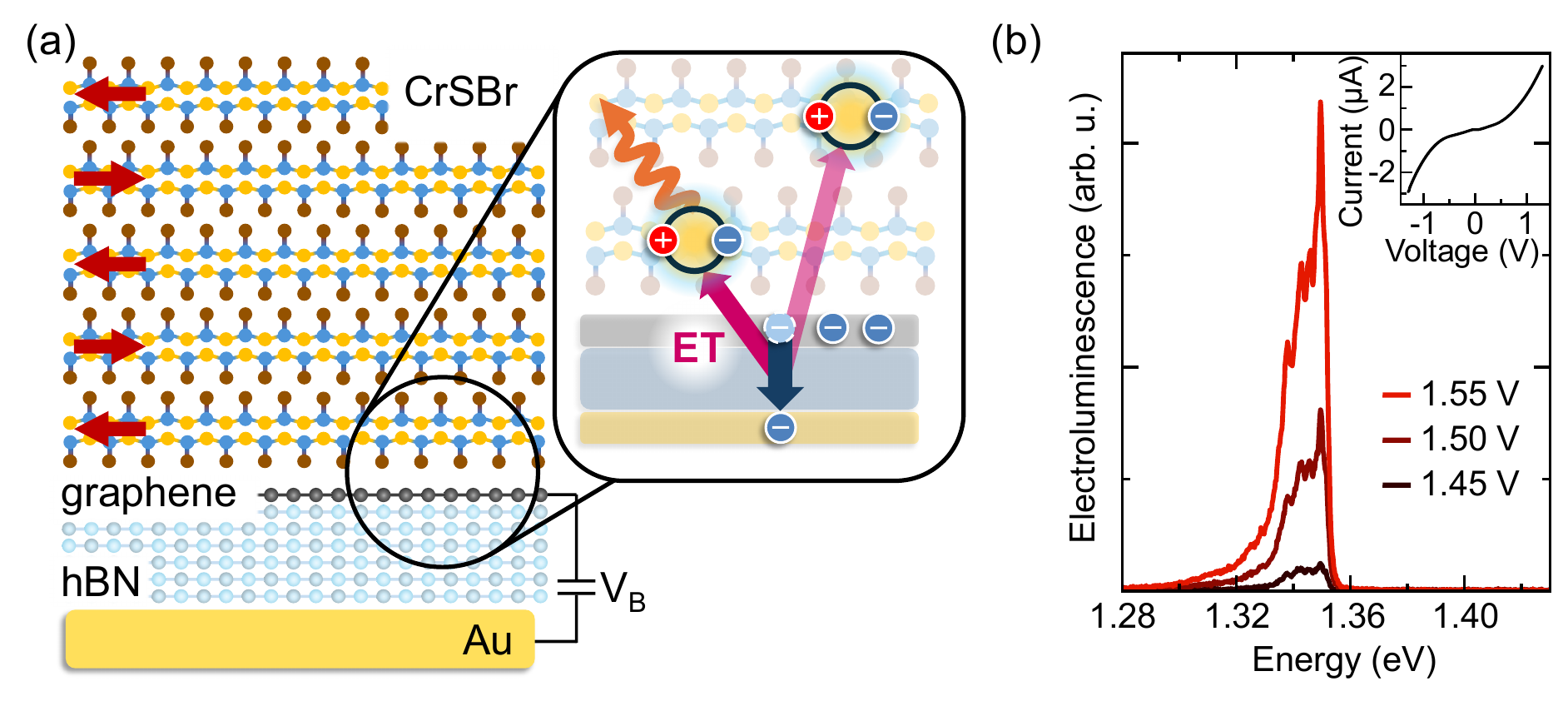}
		\caption{\textbf{Energy transfer excitation in CrSBr}.
		\textbf{a}. Schematic illustration of the device featuring a 2-250\,nm thick CrSBr layer on top of a Graphene-hBN-gold tunneling junction. Red arrows indicate the AFM order of CrSBr below the Néel temperature. Zoom-in depicts the energy transfer from a tunneling electron to excitons in the proximate lower layers of the CrSBr.
		\textbf{b}. Electroluminescence from a 25\,nm thick CrSBr flake for increasing voltage. Inset shows the corresponding tunneling current as function of the applied voltage.
		}
	\label{fig1}
\end{figure*} 

The mechanism underlying the electrical excitation, based on near-field optical processes, is illustrated in the inset of Figure 1(a). When a voltage is applied between the gold and graphene electrodes, electrons tunnel from graphene into the gold electrode. This tunneling requires conservation of both energy and momentum \cite{Parzefall2017}. Electrons undergoing elastic tunneling enter high-energy states in gold and subsequently dissipate energy through interactions with phonons within the metal. Conversely, electrons undergoing inelastic tunneling can directly emit photons, generating broadband radiation corresponding to their energy loss \cite{Parzefall2015,Kern2015, Qian2018}. When a strong dipolar excitation is nearby, tunneling electrons can alternatively transfer their energy directly to this excitation \cite{Foerster1948}. The efficiency mainly depends on three factors: distance, oscillator strength of the exciton and its dipole orientation. Here, the exceptionally strong exciton oscillator strength of CrSBr and the van der Waals nature make it a highly promising material for this excitation mechanism.

In this scenario, tunneling electrons couple to excitons in adjacent layers, thereby directly exciting excitons in the neighboring material. This method offers distinct advantages: excitons can be excited without applying an electric field directly across the active material, enabling the study of intrinsic excitonic properties and avoiding degradation in sensitive samples. Moreover, direct exciton generation prevents charge carrier imbalance. Lastly, this "open electrode" - design excites excitons irrespective of the CrSBr thickness, as the CrSBr layer is placed on top of the tunnel junction.

The devices are fabricated using a dry-pick up method \cite{Zomer2014}, stacking mechanically exfoliated CrSBr, graphene and thin (4-7 layer) tunneling hBN. All fabrication steps for handling CrSBr are performed inside an Argon-filled glovebox to prevent oxidation and ensure clean interfaces. Devices are further capped with a hBN layer to protect the devices from the environment for the measurement \cite{Telford2022,Torres2023}. The final stack is released on prepatterned gold contacts, which serve both as the bottom tunneling layer and the contact for the graphene layer (see methods for more detail). For electroluminescence measurements, we apply a voltage between the graphene and the bottom gold layer across the thin tunneling hBN.
An exemplary electroluminescence spectrum acquired at cryogenic temperatures is shown in Figure 1(b) for a 25\,nm thick CrSBr flake on top of a tunneling junction. The emission spectrum is dominated by a single resonance at around 1.35\,eV, with additional smaller resonances at lower energies. The electroluminescence intensity strongly increases with increasing voltage. Similarly, the corresponding IV-curve (see inset of Figure 1(b)) exhibits a clear tunneling behavior, with an exponentially increasing current at high voltages. 

\subsection{Electrically excited excitons in CrSBr}
An exemplary device is shown in the optical microscope image in Figure 2(a). The open electrode design enables the use of arbitrary thickness of the CrSBr, the one in the image is around 25\,nm (around 30 layers) thick. It covers part of the graphene-hBN-gold tunnel junction on the right side. The EL of this device is shown in the top part of Figure 2(b) together with the respective PL spectrum (black). The electrically excited emission (red) is dominated by a main resonance at an energy of 1.35\,eV with additional peaks at lower energy. The periodicity of the additional features might indicate an origin related to phonon replicas. Different spacings have been reported in literature, whereas the observed spacing is on the lower order \cite{Lin2024}.

\begin{figure*}[h]
	\centering
			\includegraphics[scale=0.47]{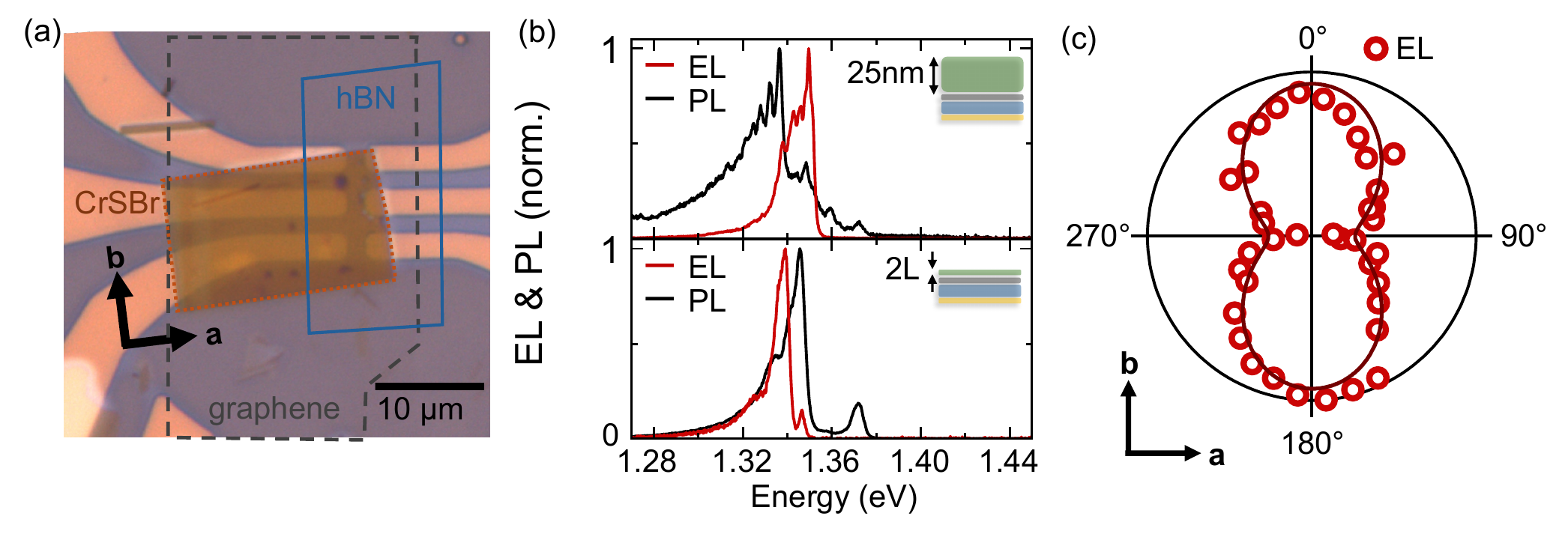}
		\caption{\textbf{Electrically excited excitons in CrSBr}.
        \textbf{a}. Optical microscope image of a tunneling junction with 25\,nm thick CrSBr layer on top. Top left shows additional flake transfered during stacking, with the typical needle-like shape indicating the crystallographic a and b axis.
        \textbf{b}. Comparison of EL and PL for a thick flake (25\,nm, top panel) and a bilayer (bottom panel) CrSBr. 
		\textbf{c}. Linear polarization of the EL as polar plot, with a and b axis indicated. Dark red line is a cos$^2$ fit to the data.
		}
	\label{fig2}
\end{figure*} 

Interestingly, the main peak of the PL is red-shifted by around 15\,meV compared to the EL, while exhibiting similar additional small features. This shift might be due to different intensity distributions, as the PL exhibits a distinct peak at the same energy, but with less intensity. The change in intensity distribution can be  attributed to the specific excitation mechanism: For PL, excitons are excited homogenously throughout the sample by a red laser, while for EL only very few layers in close proximity to the tunneling junction are excited by the near-field excitation. This changes the overall coupling to different states and modes in the system. 

However, instead of a change in intensity distribution, the optically and electrically excited emission can originate from different states. The lowest-lying layers experience a different dielectric environment because of the graphene layer, leading to screened Coulomb interactions and a change in exciton energy \cite{Liebich2025}. In addition to the dielectric surrounding, the magnetic domains also change drastically at the interface. Recently, surface excitons have been reported in CrSBr few-layers, which exhibit magnetic confinement to the surface layer \cite{Shao2025}. Due to the near-field nature of the electrical excitation, the surface excitons would be the main excitation of our electric devices. As the contribution from surface excitons increases with decreasing layer number, we further investigate this in a bilayer device, shown in the lower panel of Figure 2(b). Here, the PL exhibits both bulk exciton at 1.37\,eV and the surface exciton at around 1.34\,eV. In the bilayer, the EL is indeed more similar to the PL and mainly consists of one resonance slightly below 1.34\,eV, which we attribute to the strong surface exciton. The surface state is missing in the EL, which might be related to the additional distance from the electrode. 

Moreover, we observe a strong polarization anisotropy of the EL, as shown in the polarization-resolved measurement in Figure 2(b). This is a clear hallmark of excitons in CrSBr, where the quasi-1D bandstructure leads to a strong anisotropy of the optical resonance \cite{Klein2023a, Wilson2021}. Consequently, the high-intensity axis aligns well with the b-axis of the crystal. This gives further evidence of the electrical excitation of excitons in CrSBr flakes ranging from bilayer to bulk. It is relevant to point out that we do not observe evidence for charged exciton states, independent of sample thickness and bias polarity \cite{TabatabaVakili2024}. The trion states would appear around 20\,meV below the main exciton resonance, while we observe a blue-shifted EL compared to PL. This shows that the near-field energy transfer directly excites charge-neutral excitons, contrasting conventional charge carrier injection mechanisms.

\subsection{Self-hybridized exciton polaritons in bulk CrSBr}
We now turn to the emission of a 104\,nm thick layer to gain further insight into the coupling to optical modes of the CrSBr slab. In Figure 3(a) we show the EL together with the PL and the derivative of the reflectance contrast dR/dE (blue). EL and PL are slightly shifted while exhibiting the same shape, which consists of multiple peaks. This is distinctly different from the bilayer case, and already observable in the PL of the 25\,nm thick sample. Importantly, each of the EL peaks exhibits a corresponding resonance in the reflectance contrast. This is a clear signature for self-hybridized polariton states recently reported, where each peak is one polariton branch \cite{Dirnberger2023,Canales2021}. Here, the bare crystals confine photons due to the drastic change in refractive index at the interface with air on one side and the gold electrode on the other side. 
\begin{figure*}[h]
	\centering
			\includegraphics[scale=0.5]{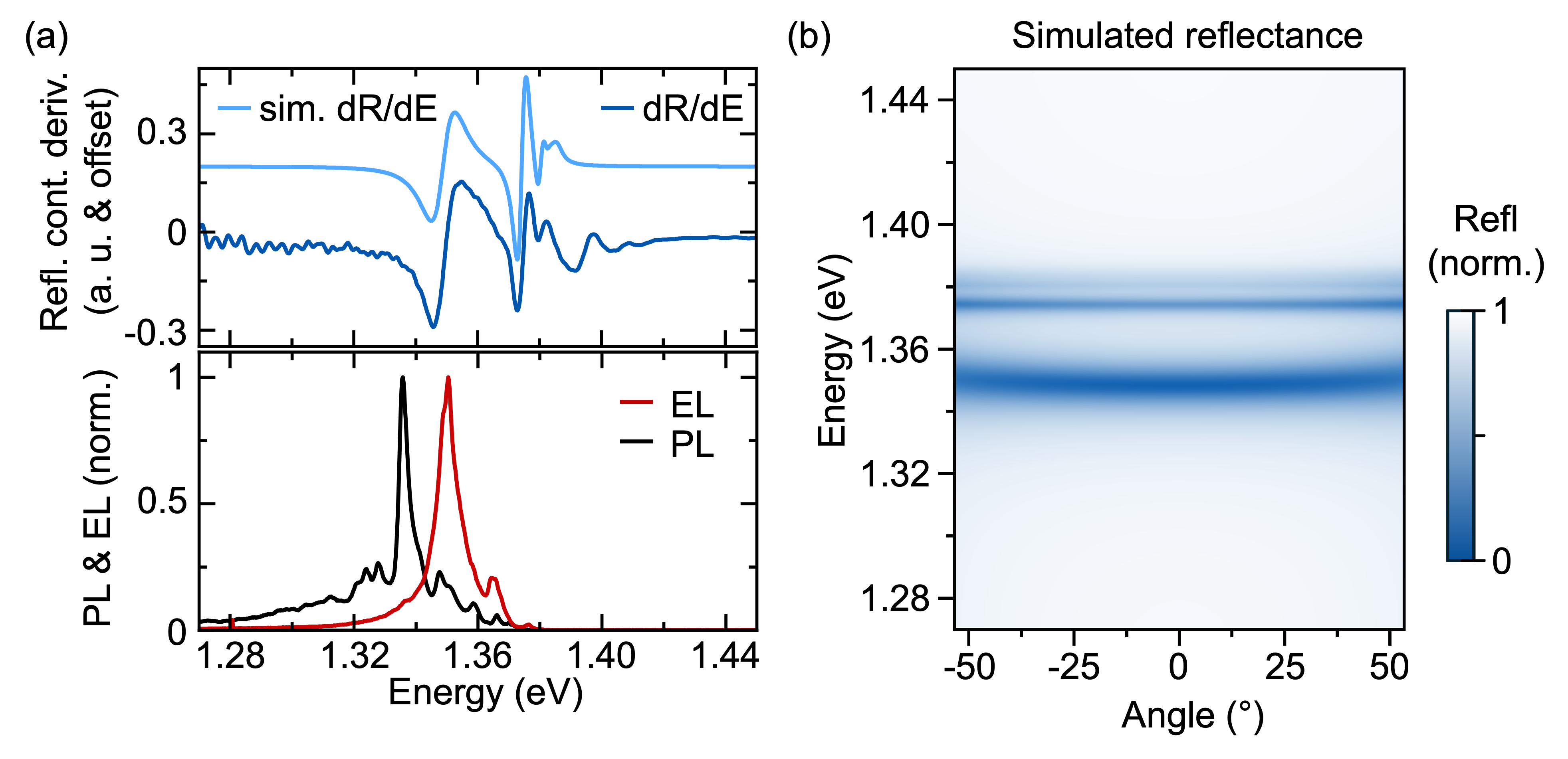}
		\caption{\textbf{Self-hybridized exciton polaritons in thick CrSBr.}
		\textbf{a}. Comparison of the reflectance (top panel) with the emission (bottom panel) of a 104\,nm thick flake.  Reflectance contrast derivative is depicted in dark blue, corresponding simulated response in light blue. Main resonances of electroluminescence (red) and photoluminescence (dark blue) are well matched with the absorption-type measurement. 
		\textbf{b}. Angle resolved simulation of the reflectance for a sample matching the experimental device, where we assume a single exciton resonance at 1.387\,eV. The additional resonances are due to the coupling between the light modes inside the CrSBr flake and the strong exciton resonance. Simulation in (a) is the derivative of the angle-integrated spectrum.
        }
	\label{fig3}
\end{figure*} 

The individual modes are well captured by the simulated reflectance in Figure 3(b), which is shown as function of incident angle of the light. We model the optical response of the sample by taking all the layers into account in a transfer matrix model, where we estimate the dielectric function of CrSBr with a single exciton resonance. The angle-dependent bending of the resonances highlight the polaritonic nature, whereas the overall small effect emphasizes the strong excitonic part of the observed polaritons.

\subsection{Coupling of inelastic electron tunneling}
 Having established the electrical excitation of exciton-polaritons via a near-field energy transfer, we will now discuss the efficiency of this process. The electrically excited emission as a function of voltage is shown in Figure 4(a). Here, multiple resonances are observed around 1.36\,eV, while the overall emission intensity increases strongly. We also observe weak, broadband emission at lower energies, which we attribute to the direct coupling of tunneling electrons to photons. The coupling to excitons is expected to be much more efficient than the coupling to photons due to the relaxed momentum conservation. Here, excitons can be excited with high momentum compared to photons limited to the light cone. This is directly observable in the strong exciton emission of Figure 4(a). 

\begin{figure*}[!ht]
	\centering
			\includegraphics[scale=0.5]{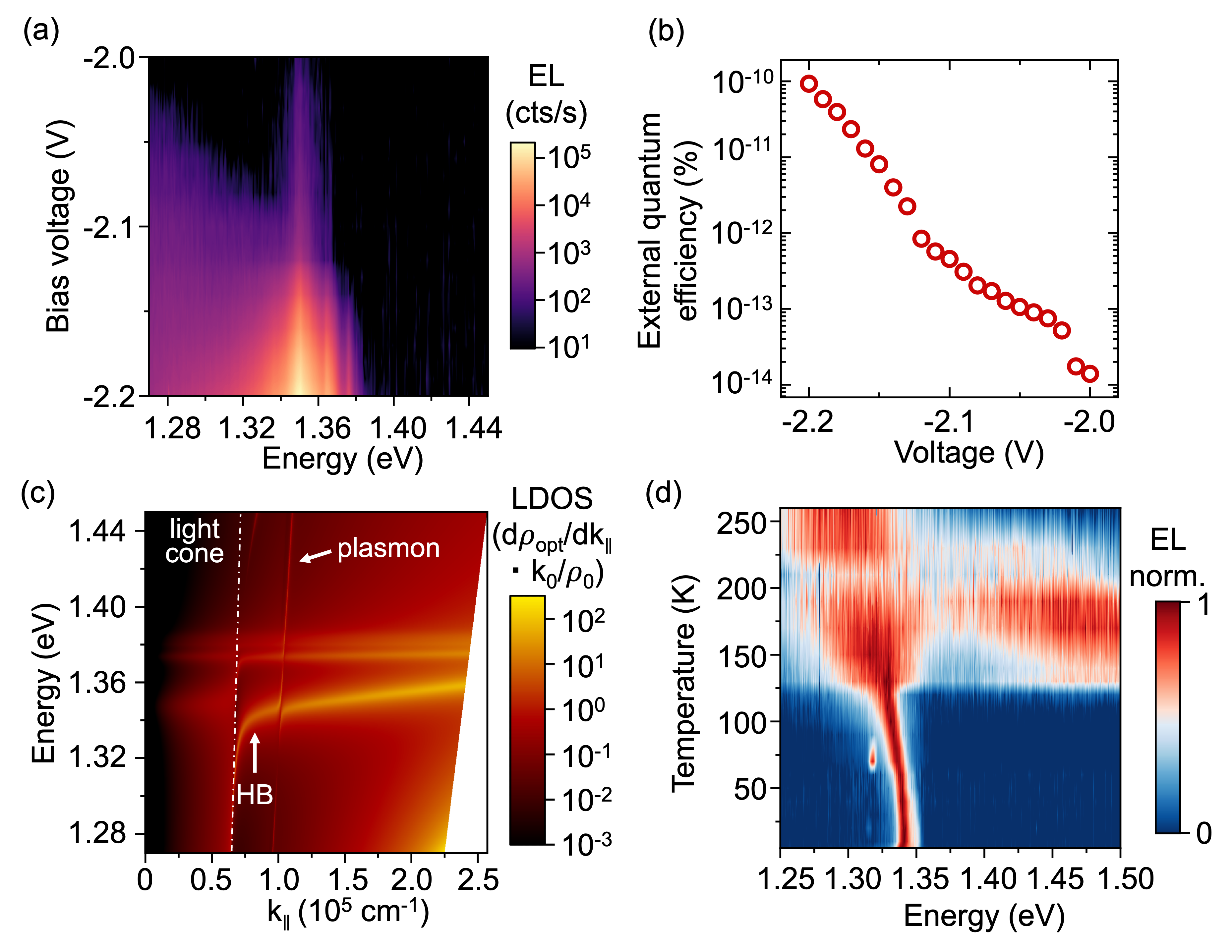}
		\caption{\textbf{Voltage- and temperature dependence of electroluminescence}
        \textbf{a}. Voltage dependent electroluminescence of a 104\,nm thick CrSBr on top of a graphene-hBN-gold junction. 
        \textbf{b}. External quantum efficiency (electron-to-photon conversion efficiency) as function of the applied voltage.
        \textbf{c}. Angular spectral density of the LDOS for an out-of-plane dipole placed 2\,nm away from a 104\,nm thick CrSBr flake (as schematically depicted in inset of Figure 1(a)). The dash-dotted line with anticrossing features close to $k_\parallel= 0.75 \cdot\,10^5 cm^{-1}$ resembles the light cone of the material. Additionally, the feature at around $k_\parallel= 1 \cdot \,10^5 cm^{-1}$ is assigned to the plasmon resonance between graphene and the gold bottom electrode. Inbetween these features, a highly dispersive mode is characterized as the hyperbolic exciton polariton mode (HB) \cite{Ruta2023}.
		\textbf{d}. Temperature dependence of the electroluminescence signal from 5\,K to 250\,K, normalized for each temperature. 
        }
	\label{fig4}
\end{figure*} 

The external quantum efficiency (EQE), shown in Figure 4(b), is overall low but exhibits a drastic increase as the applied voltage rises. Here, the presented EQE is a lower bound on the efficiency, as the CrSBr covers only partially the tunneling area. This trend is primarily attributed to the relative enhancement of inelastic tunneling versus elastic tunneling at higher voltages, which facilitates greater electron-exciton coupling \cite{Kuzmina2021}. For the investigated device we also observed EL only at elevated voltages, but in principle we would expect to see emission even below the bandgap \cite{Shan2023}. Nevertheless, the observed efficiency remains limited due to notable quenching induced by graphene, which likely varies with the Fermi level of graphene \cite{Froehlicher2018,Dong2023}. The complex interface between CrSBr and graphene also introduces variability in the charge and energy transfer dynamics \cite{Rizzo2025}. Furthermore, the efficiency can be impacted by vertical exciton transport following initial excitation of the topmost layers. Comprehensive investigations into exciton transport mechanisms in CrSBr will be required to clarify these processes.

Exciton generation is tied to the inelastic tunneling rate \cite{Papadopoulos2022,Parzefall2017}, which depends on the local density of optical states (LDOS), plotted in Figure 4(c). In our samples, the LDOS is significantly enhanced due to the high oscillator strength and the formation of self-hybridized exciton-polaritons. This enhancement enables tunneling electrons to couple efficiently with excitonic states, notably even at elevated electron momenta. Additionally, the calculated LDOS for the 104 nm-thick flake indicates the presence of hyperbolic exciton polaritons \cite{Ruta2023}. These polaritons can further contribute to the observed emission via scattering at the gold edges, thus enriching the overall emission characteristics.
Furthermore, the different modes excited by the electrical excitation can influence the lineshape, as modes with varying energy are excited (see SI).

It is crucial to note that although we expect plasmon excitation originating from the gold, identified by the pronounced, nearly vertical dispersion line around $k_\parallel= 1 \cdot \,10^5 cm^{-1}$, this plasmonic excitation is not fundamentally required for the observed electroluminescence (EL). Indeed, similar EL behavior from exciton polariton states is observed in graphene-graphene tunneling devices entirely without gold in the emission area (see SI).

Exploring the temperature dependence, Figure 4(d) distinctly reveals a pronounced shift in emission near the Néel temperature of approximately 130\,K. This transition is accurately described by a simultaneous decrease in oscillator strength and an increased linewidth at the critical temperature \cite{Shao2025}. Above this transition temperature, direct exciton emission notably diminishes, and broadband emission emerges from direct electron-light coupling processes. This shift is accompanied by reduced exciton absorption following the loss of magnetic ordering. Observing emission at room temperature is a first step towards the practical relevance and applicability of these materials for future optoelectronic devices. This capability could greatly facilitate their integration into realistic device architectures beyond CrSBr, paving the way for versatile technological applications.

\section{Conclusion}
In summary, we show for the first time electrical excitation of excitons in a van der Waals magnet with variable layer thickness from cryogenic up to room temperature. For thicker crystals, self-hybridized exciton polariton states are the main excitation modes, highlighting the strong light-matter interaction in CrSBr. We establish energy transfer from tunneling electrons to excitons as an alternative to conventional charge injection, thereby introducing a new strategy for device design. This excitation mechanism als enables the in-situ study of interfaces with other materials \cite{Yang2024, Beer2024, Huo2025, Rizzo2025}, without the need of external optical excitation. Especially the expected excitation of hyperbolic exciton polaritons in CrSBr and the surface plasmon polariton at the interface with graphene offer promising research directions. Lastly, the strong coupling between excitons and the magnetic order makes electrical excitation an important tool for future applications. Our work opens new avenues toward direct read-out of magnetic states, paving the way for simpler and more efficient device architectures.

\section{Methods}

\subsection{Device fabrication}
High-quality hBN flakes (NIMS, Japan) and commercially available graphite (NGS trading) are exfoliated on Si substrates with a 100\,nm SiO$_2$ layer, which has been treated with oxygen plasma. The thickness of the tunneling hBN is verified with an AFM. CrSBr (Prague) is exfoliated on Si substrates with a 100\,nm SiO$_2$ layer without plasma treatment, as especially thin CrSBr are challenging to be transferred. The flake thickness of CrSBr is identified first by contrast, and verified with an AFM measurement of the finished device. A thin layer of polycarbonate (PC) is placed on a polydimethylsiloxane (PDMS) stamp for the dry pick-up. First, a thick (15-40\,nm) hBN layer is picked up as top layer, which is used to subsequently pick-up the CrSBr, graphene and tunneling hbN. Finally, the PC layer with the stack is released onto pre-patterned gold electrodes (50\,nm gold with a 5\,nm Chromium layer for better adhesion). Lastly, the PC film is dissolved in chloroform and isopropanol. For the 250\,nm thick flake, a tunnel junction made only from graphene, tunneling hBN and another graphene layer is stacked first as described earlier and tested. Then, the thick CrSBr is exfoliated on a PDMS layer and transferred on top of the tunnel junction.

\subsection{Electrical and optical measurements}
All electrical and optical measurements are performed inside a closed-cycle cryostat (Attodry 800). Electrical measurements are performed using a amperemeter (Keithley 6482), which is also used as bias source. For photoluminescence measurements, the sample is excited with a continuous wave helium-neon laser with a photon energy of 1.96\,eV. Emitted light is detected either directly by a EMCCD camera (Andor iXon 897) or spectrally dispersed by a spectrometer (Andor SR303) and detected by a CCD camera (Andor iDus 416). For linearly-polarized measurements, the sample is driven by same bias voltage, while a linear polarizer placed in the detection path is rotated.
The external quantum efficiency of the sample is defined as the number of photons emitted per number of injected electrons, for which transmission spectrum measurements are performed to calibrate the absolute collection efficiency of the optical setup. A halogen calibration light source with calibrated spectrum (OceanOptics HL-2000) is used to obtain the shape of the transmission spectrum. Then, the same excitation laser is used to get the absolute value of the transmission at photon energy of 1.96\,eV, which corrects the transmission spectrum obtained by the halogen light source to an absolute position. From this, we can estimate the absolute number of emitted photons during the exposure time, which is divided by the total measured current through the junction. Considering the finite slit-width of the spectrometer and the non-uniform angular photon emission of the sample and the objective with NA = 0.81, our estimation yields the lower bound of the external quantum efficiency.  

\subsection{Simulations}
Angle-resolved reflectance spectra were calculated using a standard transfer-matrix method implemented in MATLAB~\cite{benisty_method_1998}. In the simulation, a layered structure (see supplementary information) is interfaced with gold and SiO$_2$ from one side and air from the other side. The layer thicknesses were: 50\,nm gold, 2\,nm hBN, 109\,nm CrSBr, 20\,nm hBN. The refractive indices were obtained from \cite{Dirnberger2023}.
The local density of optical states $\rho_{opt}$, relativ to the vacuum density of states $\rho_0$, is calculated from the radiated power of a dipole $P$, according to \cite{Novotny2012}: $\frac{\rho_{opt}}{\rho_0} = \frac{P}{P_0}$, with $P_0$ being the power dissipated for a point dipole in vacuum. The dissipated power $P$ is calculated from: $P=\frac{1}{2}\omega \,\mathbf{p} \,Im\{\mathbf{E}(\mathbf{r_0})\}$, with the emitted dipole angular frequency $\omega$ and the dipole moment \textbf{p}. The dipole origin $\mathbf{r_0}$ is placed at the center of the tunneling hBN, and wave equations are solved to calculated the electric field \textbf{\textit{E}}.
\\


\begin{acknowledgement}
The authors thank S. Shan for fruitful discussions and assistance in sample fabrication.
This work was supported by the Swiss National Science Foundation (grant 200020\_192362/1) and the ETH Grant SYNEMA ETH-15 19-1. J.D.Z. gratefully acknowledges financial support by the ETH Zürich Postdoctoral Fellowship programme. F.D. gratefully acknowledges support by the Emmy Noether Program (Project-ID 534078167). K.W. and T.T. acknowledge support from the JSPS KAKENHI (Grant Numbers 21H05233 and 23H02052), the CREST (JPMJCR24A5), JST and World Premier International Research Center Initiative (WPI), MEXT, Japan. Z.S. was supported by ERC-CZ program (project LL2101) from Ministry of Education Youth and Sports (MEYS) and used large infrastructure from MEYS project reg. No. CZ.02.1.01/0.0/0.0/15\_003/0000444 financed by the ERDF. 
\end{acknowledgement}

\bibliography{Bibliography}

\end{document}